\documentclass{PoS}

\newcommand{\third}{\mbox{\small $\frac{1}{3}$}}         
\newcommand{\R}{\mbox{\tiny $R$}}                        
\newcommand{\Si}{\mbox{\tiny $S$}}                       
\newcommand{\NS}{\mbox{\tiny $N\!S$}}                    

\def\lsim{\mathrel{\rlap{\lower4pt\hbox{\hskip1pt$\sim$}}
    \raise1pt\hbox{$<$}}}                
\def\gsim{\mathrel{\rlap{\lower4pt\hbox{\hskip1pt$\sim$}}
    \raise1pt\hbox{$>$}}}                

\title{
\vspace*{-1.25cm}
\begin{minipage}{\textwidth}
\begin{flushright}
\texttt{\footnotesize
PoS(Lattice 2011)158 \\%
DESY 11-169          \\%
Edinburgh 2011/34    \\%
Liverpool LTH 931    \\%
}
\end{flushright}
\end{minipage}\\[15pt]
\vspace*{+1.25cm}
       Nucleon sigma terms for $2+1$ quark flavours}

\ShortTitle{Nucleon sigma terms $\ldots$} 

\author{\speaker{R. Horsley}$^{\,a}$,
        Y. Nakamura$^b$,
        H. Perlt$^c$,
        D. Pleiter$^{de}$, 
        P.~E.~L. Rakow$^f$,
        G. Schierholz$^{eh}$,
        A. Schiller$^c$, 
        H. St\"uben$^i$, 
        F. Winter$^a$
        and J.~M. Zanotti$^a$ \\
        \llap{$^a$} School of Physics and Astronomy,
                    University of Edinburgh,
                    Edinburgh EH9 3JZ, UK \\
        \llap{$^b$} RIKEN Advanced Institute for Computational Science,
                    Kobe, Hyogo 650-0047, Japan \\
        \llap{$^c$} Institut f\"ur Theoretische Physik,
                    Universit\"at Leipzig, 04109 Leipzig, Germany \\
        \llap{$^d$} JSC, J\"ulich Research Centre,
                    52425 J\"ulich, Germany \\
        \llap{$^e$} Institut f\"ur Theoretische Physik,
                    Universit\"at Regensburg, 93040 Regensburg, Germany \\
        \llap{$^f$} Theoretical Physics Division,
                    Department of Mathematical Sciences,
                    University of Liverpool,
                    Liverpool L69 3BX, UK \\
        \llap{$^h$} Deutsches Elektronen-Synchrotron DESY,
                    22603 Hamburg, Germany \\
        \llap{$^i$} Konrad-Zuse-Zentrum f\"ur Informationstechnik Berlin,
                    14195 Berlin, Germany \\
        E-mail: \email{rhorsley@ph.ed.ac.uk} }

\author{QCDSF--UKQCD Collaboration}

\abstract{
   QCD lattice simulations yield hadron masses as functions
   of the quark masses. From the gradients of the hadron masses
   the sigma terms can then be determined. We consider here
   dynamical 2+1 flavour simulations, in which we start from a point
   of the flavour symmetric line and then keep the singlet
   or average quark mass fixed as we approach the physical point.
   This leads to highly constrained fits for hadron masses in a multiplet.
   The gradient of this path for a hadron mass then gives a relation
   between the light and strange sigma terms. A further relation
   can be found from the change in the singlet quark mass along
   the flavour symmetric line. This enables light and strange sigma terms
   to be estimated for the baryon octet.}

\FullConference{The XXIX International Symposium on Lattice Field Theory 
                - Lattice 2011 \\
                July 10-16, 2011 \\
                Squaw Valley, Lake Tahoe, California}

\begin{document}


\section{Introduction} 


In this talk we shall describe a method for the determination
of the hyperon sigma terms based on the results of \cite{horsley11a} to
which we refer to for more details including numerical results.

Sigma terms, $\sigma_l^{(H)}$, $\sigma_s^{(H)}$ are defined
as that part of the mass of the hadron (for example the nucleon)
coming from the vacuum connected expectation value of the up ($u$)
down ($d$) and strange ($s$) quark mass terms in the QCD Hamiltonian,
\begin{eqnarray}
   \sigma_l^{(H)} = m_l^{\R}\langle H|(\overline{u}u + \overline{d}d)^{\R} 
                                   |H\rangle \,,
   \qquad
   \sigma_s^{(H)} = m_s^{\R}\langle H|(\overline{s}s)^{\R} 
                                   |H\rangle \,,
\label{sig_def}
\end{eqnarray}
where we have taken the $u$ and $d$ quarks to be mass degenerate,
$m_u = m_d \equiv m_l$. (The superscript $^{\R}$ denotes a renormalised
quantity.)
Eq.~(\ref{sig_def}) is usually
written (in particular for the nucleon) as
\begin{eqnarray}
   \sigma_l^{(N)} 
      = { m_l^{\R}\langle N|(\overline{u}u + \overline{d}d
                         - 2 \overline{s}s)^{\R} |N \rangle 
        \over 1 - y^{(N)\R} } \,, \qquad
   y^{(N)\R} 
      = { 2 \langle N| (\overline{s}s)^{\R} |N \rangle \over
          \langle N| (\overline{u}u + \overline{d}d)^{\R} |N \rangle } \,,
\label{sig_y_def}
\end{eqnarray}
(i.e.\ we consider $y^{(N)\R}$ rather than $\sigma_s^{(N)}$).
The simplest calculation, (which we will
discuss in more detail later) uses first order in $SU(3)$
flavour symmetry (octet) breaking to give
\begin{eqnarray}
   \sigma_l^{(N)}
      = { m_l^{\R} \over m_s^{\R} - m_l^{\R} }
           { M_\Xi + M_\Sigma - 2M_N \over 1 - y^{(N)\R} }
              \sim { 26 \over 1 - y^{(N)\R} } \,\mbox{MeV} \,,
\label{sigl_est}
\end{eqnarray}
and
\begin{eqnarray}
   \sigma_s^{(N)}
      = { m_s^{\R} \over m_l^{\R} } {1\over 2}y^{(N)\R} \sigma_l^{(N)}
              \sim 325 {  y^{(N)\R} \over 1 - y^{(N)\R} } \,\mbox{MeV} \,,
\label{sigs_est}
\end{eqnarray}
where $m_s^{\R}/ m_l^{\R}$ is the ratio of the strange to light quark masses,
which using the leading order PCAC formula for this ratio gives
$m_s^{\R}/m_l^{\R} = (2M_K^2 - M_\pi^2)/M_\pi^2 \sim 25$.
The Zweig rule, $\langle N| (\overline{s}s)^{\R} |N \rangle \sim 0$
would then give $\sigma_l^{(N)} \sim 26\,\mbox{MeV}$, 
$\sigma_s^{(N)} \sim 0\,\mbox{MeV}$ while any non-zero strangeness content,
$y^{(N)\R} > 0$ would increase this value of $\sigma_l^{(N)}$, $\sigma_s^{(N)}$
(and indeed due to the large coefficient, $\sigma_s^{(N)}$ quite rapidly).

Computing the sigma terms from lattice QCD has a long history
from quenched to $2$ flavour and more recently $2+1$ flavour
simulations. In general more recent results
tend to give lower values than earlier determinations.

In this talk, we shall investigate this simple picture
as described above and in particular test the linearity
assumption of $SU(3)$ flavour symmetry breaking.


\section{Flavour symmetry expansions}
\label{flavour_sym_expan}


Lattice simulations start at some point in the $(m_s^{\R}, m_l^{\R})$
plane and then approach the physical point $(m_s^{\R\,*}, m_l^{\R\,*})$
along some path. (In future we shall denote the physical point
with a $^*$.) As we shall be considering flavour symmetry breaking
then we shall start here at a point on the flavour symmetric line
$m_l^{\R} = m_s^{\R}$ and then consider the path keeping the average
quark mass constant, $\overline{m} = \mbox{const.}$. The $SU(3)$
flavour group (and quark permutation symmetry) then restricts
the quark mass polynomials that are allowed,
giving for the baryon octet
\begin{eqnarray}
   M_H = M_0(\overline{m}) + c_H\delta m_l + O(\delta m_l^2) \,,
\label{baryon_octet_linfit}
\end{eqnarray}
with
$c_N = 3A_1$, $c_\Lambda = 3A_2$, $c_\Sigma = -3A_2$, $c_\Xi = -3(A_1-A_2)$
and
\begin{eqnarray}
   \delta m_l   = m_l - \overline{m}\,, \qquad 
   \overline{m} = \third (2m_l + m_s) \,.
\end{eqnarray}
So to linear order in the quark mass, we only have two unknowns,
$A_1$, $A_2$ (rather than four). A similar situation also holds
for the pseudoscalar and vector octets (one unknown) and baryon decuplet
(also one unknown). This highly constrains the numerical fits.

Permutation invariant functions of the masses, $X_S$, (or `centre of mass'
of the multiplet) have no linear dependence on the quark mass.
For example for the baryon octet we have
\begin{eqnarray}
   X_N = \third( M_N + M_\Sigma + M_\Xi)
           = M_0(\overline{m}) +  O(\delta m_l^2) \,.
\label{XN_def}
\end{eqnarray}
(The corresponding result for the pseudoscalar octet is given
later in eq.~(\ref{pseudoscalar_octet_linfit}).)

Furthermore expanding about a specific fixed point, $m_l = m_s = m_0$
on the flavour symmetric line and allowing $\overline{m}$ to vary,
we then have
\begin{eqnarray}
   M_0(\overline{m}) = M_0(m_0) + M_0^\prime(m_0)(\overline{m}-m_0)
                                + O((\overline{m}-m_0)^2) \,.
\label{M0_expan}
\end{eqnarray}
We will see that $A_1$, $A_2$ determine all the non-singlet
sigma terms and $M^\prime(m_0)$ the singlet sigma terms. 

\begin{figure}[htb]
   \begin{minipage}{0.40\textwidth}
   \includegraphics[width=7.00cm]{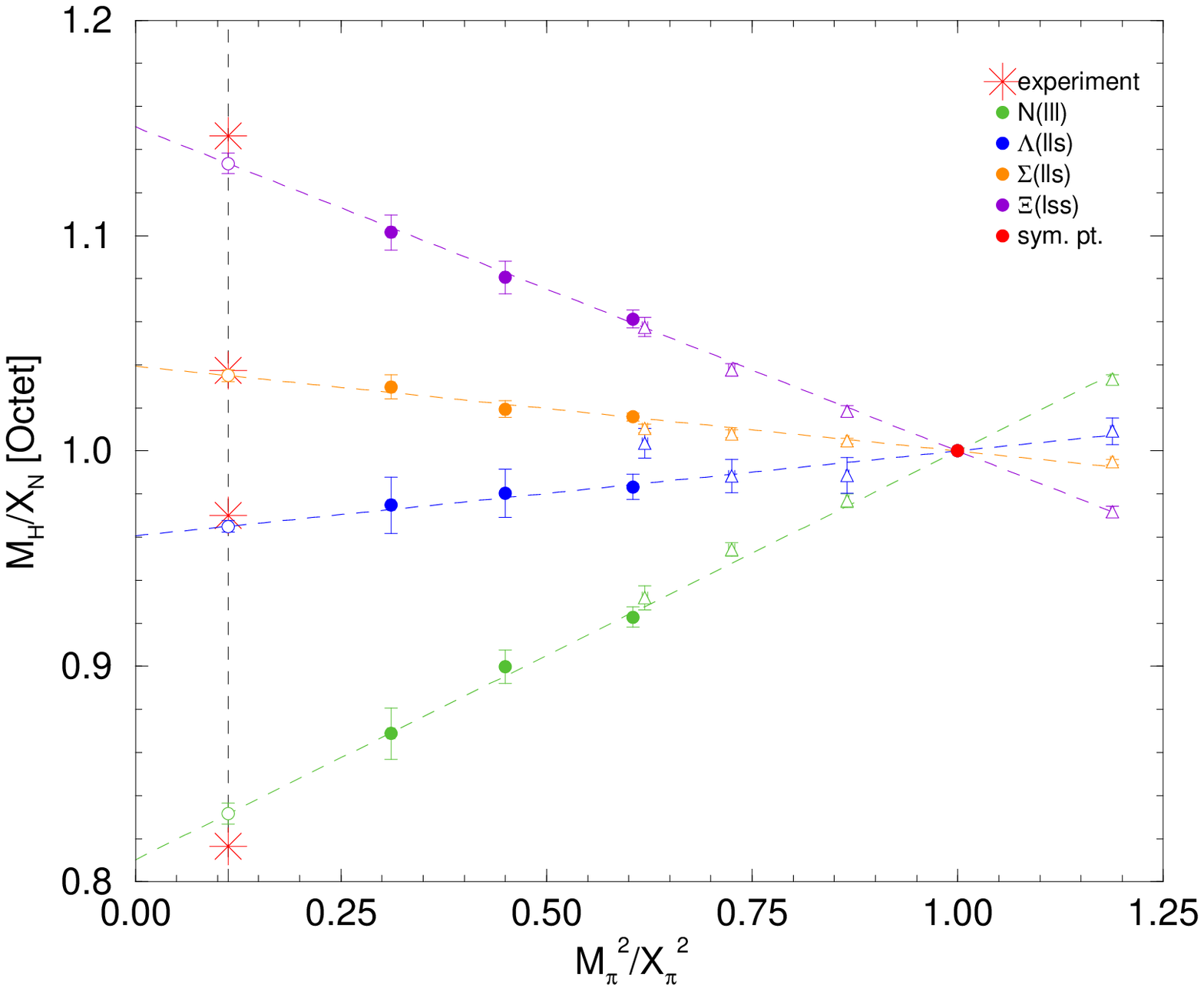}
   \end{minipage} \hspace*{0.10\textwidth}
   \begin{minipage}{0.40\textwidth}
   \includegraphics[width=7.00cm]{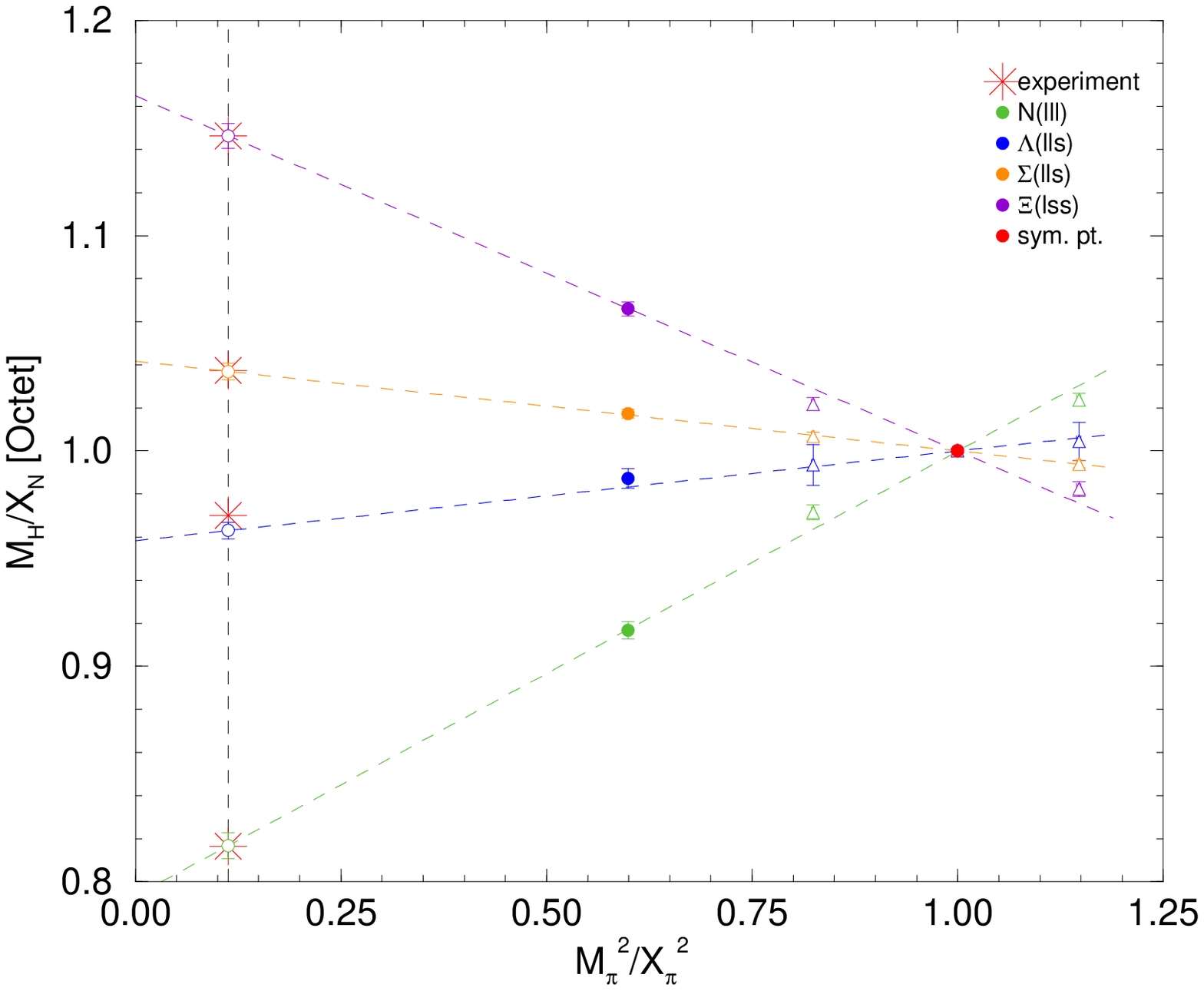}
   \end{minipage}
   \caption{$M_H/ X_N$ ($H = N$, $\Lambda$, $\Sigma$, $\Xi$)
            against $M_\pi^2/X_\pi^2$ for initial point on the
            flavour symmetric line given by $\kappa_0 = 0.12090$,
            left panel and $\kappa_0 = 0.12092$, right panel.
            The $32^3\times 64$ lattices are filled circles,
            while the $24^3\times 48$ lattices are open triangles.
            Also shown is the combined fit of
            eq.~(\protect\ref{mNoXN_mps2oXpi2}) (the dashed lines)
            to the $32^3\times 64$ lattice data. The fit results
            are the open circles, while the experimental points are
            the (red) stars.}
\label{mpsO2omNOpmSigOpmXiOo32_mNOomNOpmSigOpmXiOo3_32x64_lin}
\end{figure}

As an example of the quark mass expansion from a point on the flavour
symmetric line in
Fig.~\ref{mpsO2omNOpmSigOpmXiOo32_mNOomNOpmSigOpmXiOo3_32x64_lin}
we plot the baryon octet $M_H/ X_N$ for $H = N$, $\Lambda$,
$\Sigma$, $\Xi$ against $M_\pi^2/X_\pi^2$ together with a linear
fit, eq.~(\ref{baryon_octet_linfit}) and implicitly
eq.~(\ref{pseudoscalar_octet_linfit}) using $2+1$ $O(a)$
improved clover fermions at $\beta = 5.50$,
using two starting values for the quark mass
on the flavour symmetric line. All the points have been arranged
in the simulation to have constant $\overline{m}$. We see that
a linear fit provides a good description of the numerical data
from the symmetric point (where $M_\pi \sim X_\pi^* = 410.9\,\mbox{MeV}$)
down to the physical pion mass.


\section{(Hyperon) $\sigma$ equations}
\label{hyperon_sigma_eqns}


\subsection{Renormalisation}


For Wilson (clover) fermions under renormalisation the singlet and
non-singlet pieces of the quark mass renormalise differently.
We have
\begin{eqnarray}
   m_q^{\R} = Z^{\NS}\left[m_q + \alpha_Z\third(2m_l+m_s)\right]\,,
             \qquad \alpha_Z = { Z^{\NS} - Z^{\Si} \over Z^{\NS}} \,.
\label{renorm_mq}
\end{eqnarray}
In the action the term
$\sum_q m_q \overline{q}q = \sum_q m_q^{\R} (\overline{q}q)^{\R}$ i.e.\ is
a renormalisation group invariant or RGI quantity. Upon writing this
in a matrix form and inverting gives
\begin{eqnarray}
   (\overline{q}q)^{\R} 
     = {1 \over Z^{\NS}}
          \left[ \overline{q}q 
                 - {\alpha_Z \over 1+\alpha_Z}
                      \third(\overline{u}u+\overline{d}d+\overline{s}s)
          \right] \,,
\label{qqbar_ren}
\end{eqnarray}
so for $\alpha_Z \not=0$ then there is always mixing between bare operators.
Useful quark combinations are the octet%
\footnote{Eq.~(\ref{ren_octet}) also leads to eq.~(\ref{sigl_est})
as from section~\ref{sigma_eqns} we have
$\langle N|(\overline{u}u+\overline{d}d-2\overline{s}s)^{\R}|N\rangle
= c_N/Z^{\NS} = 3A_1/Z^{\NS}$. Together with 
$M_\Xi + M_\Sigma -2M_N = - 9A_1\delta m_l = 3A_1(m_s^{\R} - m_l^{\R})/Z^{\NS}$ 
this gives  eq.~(\ref{sigl_est}). An alternative mass combination
that also picks out the $A_1$ coefficient is
$ M_\Xi - M_\Lambda = - 3A_1\delta m_l$.}
and singlet combinations, namely
\begin{eqnarray}
   (\overline{u}u+\overline{d}d)^{\R} - 2(\overline{s}s)^{\R}
    &=& {1 \over Z^{\NS}}\,
         \left( (\overline{u}u+\overline{d}d) - 2(\overline{s}s) \right)
                                                    \label{ren_octet}   \\
   (\overline{u}u+\overline{d}d)^{\R} + (\overline{s}s)^{\R}
    &=& {1 \over Z^{\NS}(1+\alpha_Z)}\,
         \left( (\overline{u}u+\overline{d}d) + (\overline{s}s) \right) \,.
\label{useful_comb}
\end{eqnarray}


\subsection{$\sigma$ equations}
\label{sigma_eqns}


Scalar matrix elements can be determined from the gradient of the hadron
mass (with respect to the quark mass) by using the Feynman--Hellman theorem
which is true for both bare and renormalised quantities. So if we take the
derivative with respect to the bare quark mass we get the bare
$\overline{q}q$ matrix element,
\begin{eqnarray}
   {\partial M_H \over \partial m_l } 
      = \langle H| (\overline{u}u + \overline{d}d) | H \rangle\,, \qquad
   {\partial M_H \over \partial m_s } 
      = \langle H| \overline{s}s | H \rangle \,.
\label{fh_thm}
\end{eqnarray}
Multiplying the renormalised quark mass, eq.~(\ref{renorm_mq}),
together with eqs.~(\ref{useful_comb}) (or more generally with
eq.~(\ref{qqbar_ren})) we can find RGI combinations (i.e.\ a form
where the renormalisation constant $Z^{\NS}$ cancels).
In particular we find
\begin{eqnarray}
   \sigma_l^{(H)} - 2r \sigma_s^{(H)}
      &=& {3r \over 1 + 2r}(1+\alpha_Z)m_0c_H
\label{sigl_sigs_simul}                                      \\
   \sigma_l^{(H)} + r \sigma_s^{(H)}
      &=& {3r \over 1 + 2r}m_0 M_0^\prime(m_0) \,,
\label{sigs_sigl_simul}
\end{eqnarray}
where $r$ is the ratio of quark masses $r \equiv m_l^{\R} / m_s^{\R}$.
The two simultaneous equations, which can be easily solved, give
\begin{eqnarray}
   \sigma_l^{(H)} 
     &=& {r \over 1+2r}\,\left[ (1+\alpha_Z)m_0c_H 
                              + 2m_0M^{\prime}_0(m_0) \right]
                                                             \nonumber   \\
   \sigma_s^{(H)} 
     &=& {1 \over 1+2r}\,\left[ - (1+\alpha_Z)m_0c_H 
                              + m_0M^{\prime}_0(m_0) \right] \,.
\label{sigl_sigs}
\end{eqnarray}
We see that the smallness of $\sigma_l^{(H)}$ in comparison to
$\sigma_s^{(H)}$ is certainly guaranteed by the presence of
an additional $r$ in its numerator. These coefficients are also
sufficient to determine $y^{(H)\R}$, as can be seen from
eq.~(\ref{sigl_sigs}),
\begin{eqnarray}
   y^{(H)\R} = 2\, { -(1+\alpha_Z)m_0c_H + m_0M^{\prime}_0(m_0) \over
                   (1+\alpha_Z)m_0c_H + 2m_0M^{\prime}_0(m_0) } \,.
\label{yHR_formula}
\end{eqnarray}
It is seen that $y^{(H)\R}$ only depends on gradients and not
on the physical point.

It is convenient to normalise the coefficients by $X_N$ 
so we now need to find the coefficients $(1+\alpha_Z)m_0c_H/X_N(m_0)$
and $m_0M^{\prime}_0(m_0)/X_N(m_0)$ and also to extrapolate to
the point where the quark mass ratio takes its physical value,
i.e.\ $r = r^*$.


\subsection{Determination of the coefficients}
\label{det_coeff}


The coefficients can be determined by considering gradients with respect
to a physical quantity. As in eq.~(\ref{baryon_octet_linfit})
we also have a similar expansion for the pseudoscalar octet,
\begin{eqnarray}
   M_\pi^2 = M_{0\,\pi}^2 + 2\alpha\delta m_l + O(\delta m_l^2) \,,
\label{pseudoscalar_octet_linfit}
\end{eqnarray}
(together with $M_K^2 = M_{0\,\pi}^2 - \alpha\delta m_l + O(\delta m_l^2)$,
$M_{\eta_s}^2 = M_{0\,\pi}^2 - 4\alpha\delta m_l + O(\delta m_l^2)$).
Analogously to eq.~(\ref{XN_def}) we can define a flavour singlet quantity
$X_\pi^2 = \third(2M_K^2+M_\pi^2) = M_{0\,\pi}^2 + O(\delta m_l^2)$
However, as well as eq.~(\ref{baryon_octet_linfit}), we have
the additional constraint from PCAC $M_\pi^2 = 2B_0^{\R}m_l^{\R}$
(together with $M_K^2 = B_0^{\R}(m_l^{\R}+m_s^{\R})$).
If we now consider an expansion in the pion mass then
eliminating $\delta m_l$ between eq.~(\ref{baryon_octet_linfit})
and eq.~(\ref{pseudoscalar_octet_linfit}) gives
\begin{eqnarray}
   {M_H \over X_N} 
      = \left( 1 - [(1+\alpha_Z)m_0{c_H \over X_N}] \right)
        + [(1+\alpha_Z)m_0{c_H \over X_N}]\, {M_\pi^2 \over X_\pi^2} \,,
\label{mNoXN_mps2oXpi2}
\end{eqnarray}
from the point on the symmetric line $m_0 = \overline{m}$.
Thus if we plot $M_H / X_N$ versus $M_\pi^2 / X_\pi^2$ (holding the
singlet quark mass, $\overline{m}$ constant) then the gradient
immediately yields $(1+\alpha_Z)m_0c_H/X_N$. The only assumption is that
the `fan' plot splittings remain linear in $\delta m_l$ down to
the physical point. In 
Fig.~\ref{mpsO2omNOpmSigOpmXiOo32_mNOomNOpmSigOpmXiOo3_32x64_lin}
we show this plot leading to a results for
$(1+\alpha_Z)m_0 c_H/X_N$ for $\kappa_0 = 0.12090$, $0.12092$.

\begin{figure}[htb]
   \begin{minipage}{0.40\textwidth}
   \begin{center}
      \includegraphics[width=6.25cm,height=6.25cm]
         {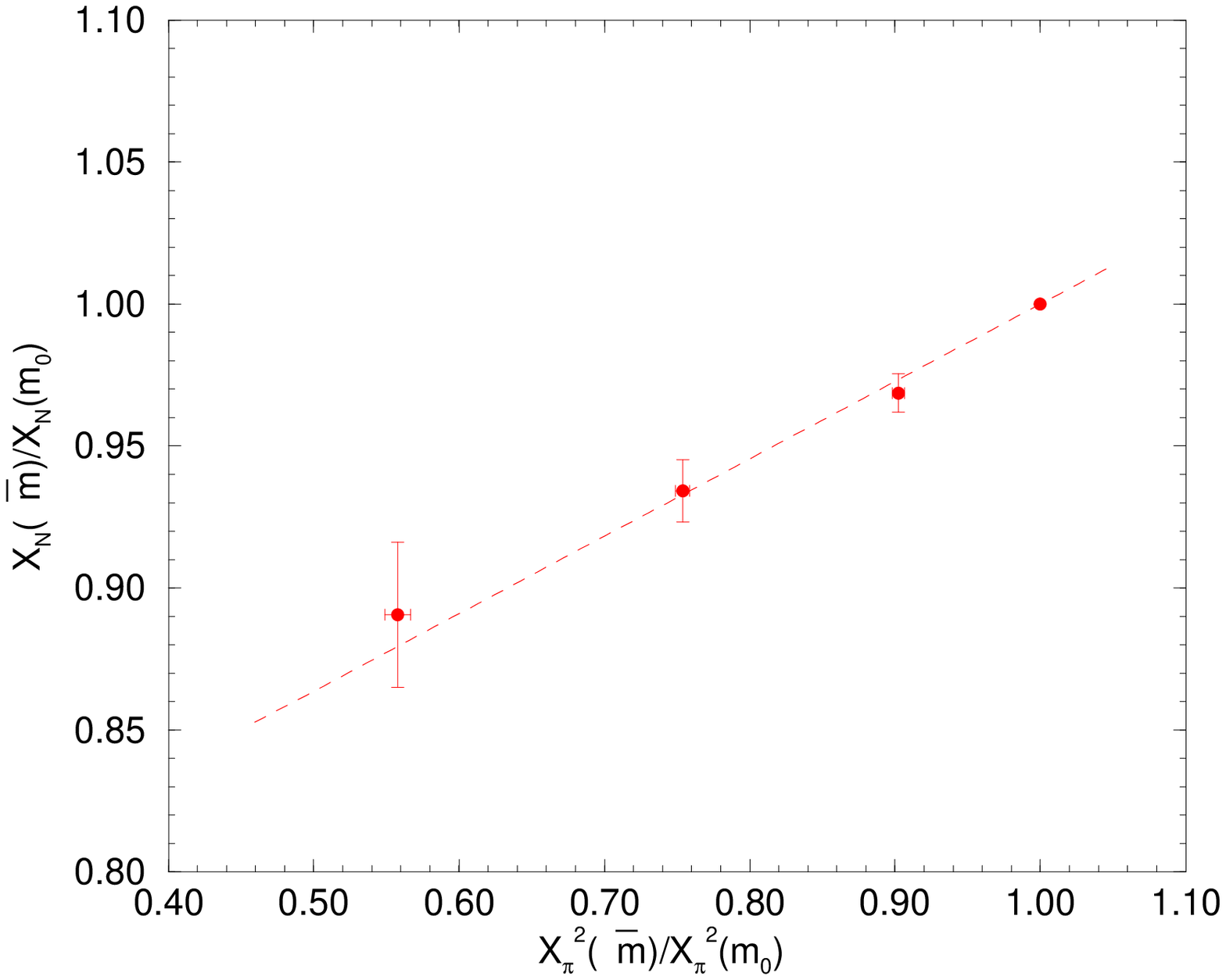}
   \end{center}
   \end{minipage} \hspace*{0.05\textwidth}
   \begin{minipage}{0.40\textwidth}
   \begin{center}
   \includegraphics[width=7.00cm]{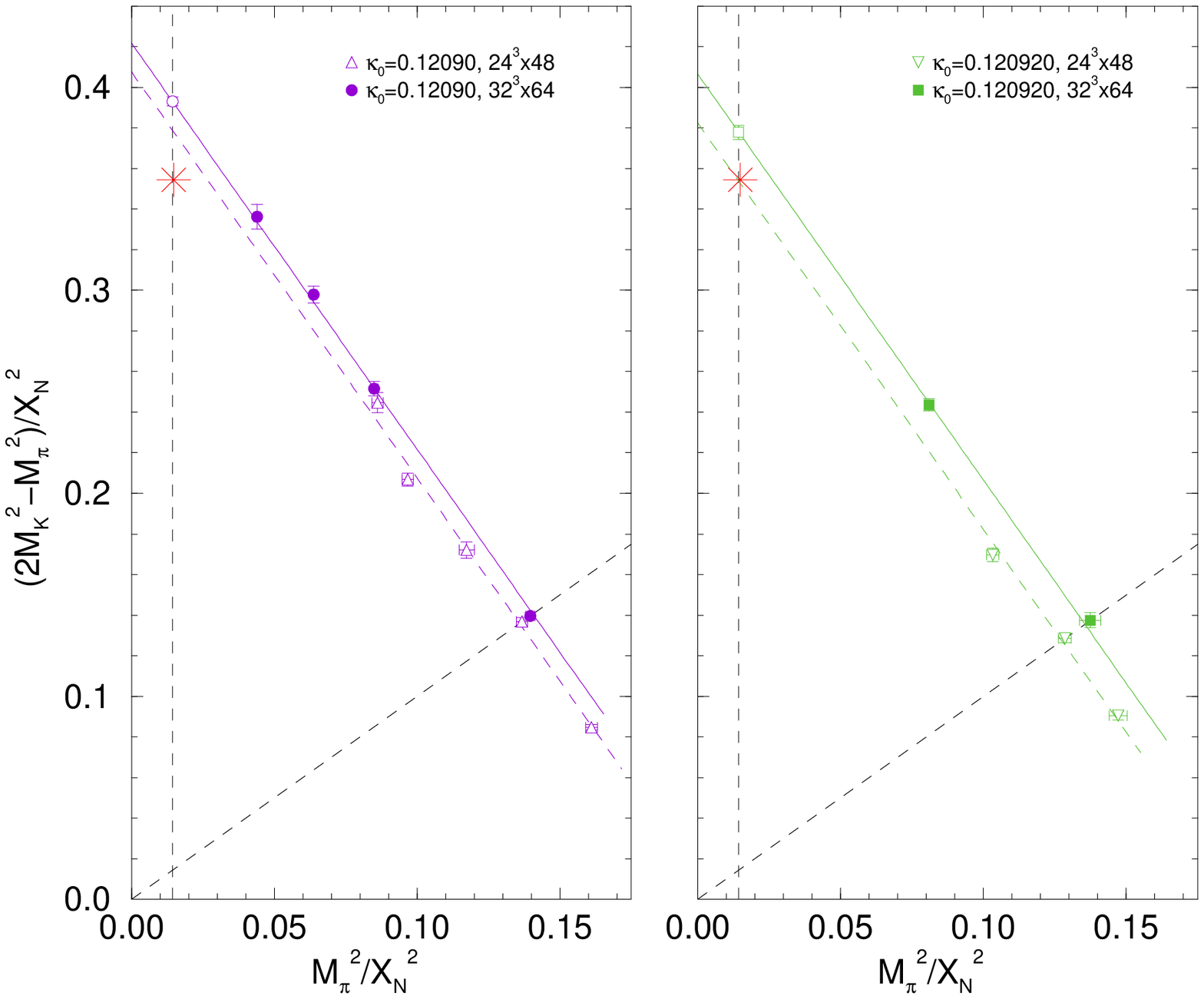}
   \end{center}
   \end{minipage}
   \caption{Left plot:$X_N(\overline{m}) / X_N(m_0)$ versus
            $X_\pi^2(\overline{m}) / X_\pi^2(m_0)$ along the
            flavour symmetric line, together with the linear fit
            from eq.~(\protect\ref{XNoXN_Xpi2oXpi2});
            Right plot: $(2M_K^2-M_\pi^2)/X_N^2$ versus $M_\pi^2 / X_N^2$
            for $\kappa_0 = 0.12090$ (left panel)
            and $\kappa_0 = 0.12092$ (right panel).
            The $32^3\times 64$ volume results are given by
            the filled symbols, while the $24^3\times 48$
            volume results are shown using empty triangles.
            The fit is given in eq.~(\protect\ref{const_mbar_fit}).
            Experimental points are denoted by (red) stars.}
   \label{b5p50_XNoXNkp12090_Xpi2oXpi2kp12090+mps2oXn2_2mpsK2-mps2oXn2}
\end{figure}

Furthermore on the flavour symmetric line eliminating
$(\overline{m} - m_0)$ between eqs.~(\ref{M0_expan})
and the corresponding one for $M_\pi^2(\overline{m})$ gives
\begin{eqnarray}
   {X_N(\overline{m}) \over X_N(m_0) }
     = \left( 1 - [{m_0 M_0^\prime(m_0) \over X_N(m_0)}] \right)
        + [{m_0 M_0^\prime(m_0) \over X_N(m_0)}] \,
           {X_\pi^2(\overline{m}) \over X_\pi^2(m_0)} \,.
\label{XNoXN_Xpi2oXpi2}
\end{eqnarray}
Again in a plot of $X_N(\overline{m}) / X_N(m_0)$ versus
$X_\pi^2(\overline{m}) / X_\pi^2(m_0)$ the gradient immediately
gives the required ratio $m_0M_0^\prime(m_0) / X_N(m_0)$. In 
Fig.~\ref{b5p50_XNoXNkp12090_Xpi2oXpi2kp12090+mps2oXn2_2mpsK2-mps2oXn2}
we plot $X_N(\overline{m}) / X_N(m_0)$ versus
$X_\pi^2(\overline{m}) / X_\pi^2(m_0)$. From eq.~(\ref{XNoXN_Xpi2oXpi2})
the gradient gives the required number.

Finally the quark mass ratio, $r$, must be estimated.
In the right panel of
Fig.~\ref{b5p50_XNoXNkp12090_Xpi2oXpi2kp12090+mps2oXn2_2mpsK2-mps2oXn2}
we plot $(2M_K^2-M_\pi^2)/X_N^2$ versus $M_\pi^2 / X_N^2$.
From eq.~(\ref{pseudoscalar_octet_linfit}) we have
\begin{eqnarray}
   {2 M_K^2 - M_\pi^2 \over X_N^2}
       = 3{M_{0\pi}^2 \over X_N^2} - 2{M_\pi^2 \over X_N^2} \,.
\label{const_mbar_fit}
\end{eqnarray}
As in section~\ref{flavour_sym_expan}, we see that for constant
$\overline{m}$ the data points lie on a straight line
(i.e.\ there is an absence of significant non-linearity).
Together with PCAC, this gives the $x$-axis is proportional to
$m_l^{\R}$ while the $y$-axis is proportional to $m_s^{\R}$ and thus the
ratio gives $r$. Taking our physical scale to be defined
from $M_\pi^2/X_N^2|^*$ (i.e.\ from the $x$-axes of
Fig.~\ref{b5p50_XNoXNkp12090_Xpi2oXpi2kp12090+mps2oXn2_2mpsK2-mps2oXn2})
gives $1/r^*$.

What can we say about corrections to the linear terms?
The simple linear fit describes the data well, from the symmetric
point to our lightest pion mass, both along the
$\overline{m} = \mbox{const.}$  line and the flavour symmetric line.
To see the possible influence of curvature we compare linear fits
with quadratic fits as discussed in the Appendix
of \cite{horsley11a}. These will be used in section~\ref{results}
for the estimate of possible systematic effects.


\section{Results and Conclusions}
\label{results}


We can now determine $y^{(H)\R}$ and $\sigma_l^{(H)}$,
$\sigma_s^{(H)}$. The scale is taken as $X_N = 1.1501\,\mbox{GeV}$.
We shall only discuss here the general details of the results;
the numerical values are given in \cite{horsley11a}.

From eq.~(\ref{sigl_sigs_simul}) we can find an indication of
the magnitude of $\sigma^{(N)}_l$ as approximately
\begin{eqnarray}
   \sigma_l^{(N)\,*} 
      \sim 22.4 + {\sigma_s^{(N)\,*} \over 13.6} \, \mbox{MeV}
      \gsim 22.4 \, \mbox{MeV} \,.
\label{sig_mbarconst_line}
\end{eqnarray}
The last inequality follows as obviously
$\sigma_s^{(N)*} > 0$. Indeed this shows that a non-zero
$\sigma_s^{(N)*} > 0$ can only add a few $\mbox{MeV}$ to this result.

\begin{figure}[htb]
   \begin{minipage}{0.40\textwidth}
   \begin{center}
   \includegraphics[width=3.50cm]{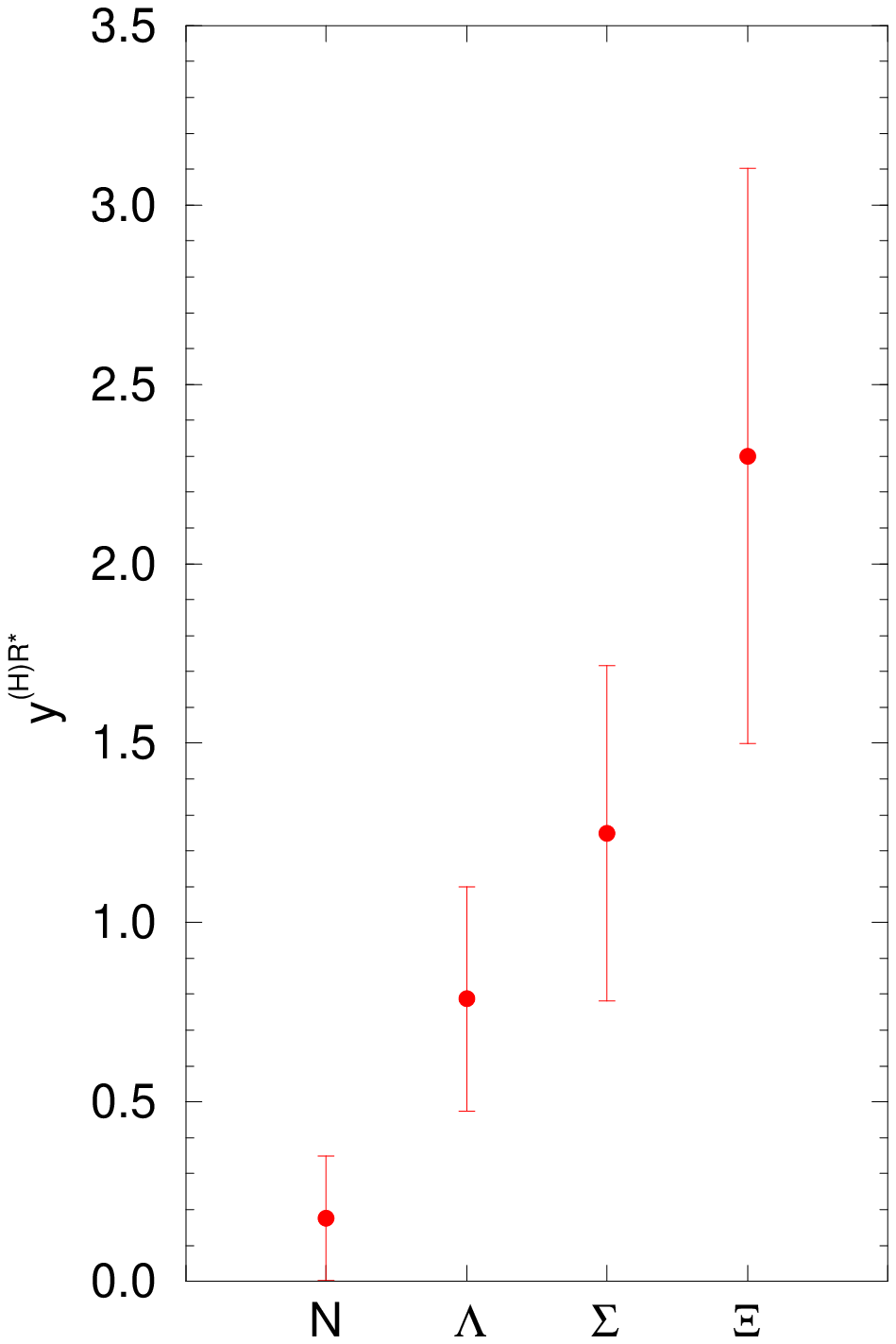}
   \end{center}
   \end{minipage} \hspace*{0.05\textwidth}
   \begin{minipage}{0.40\textwidth}
   \begin{center}
   \includegraphics[width=7.50cm]{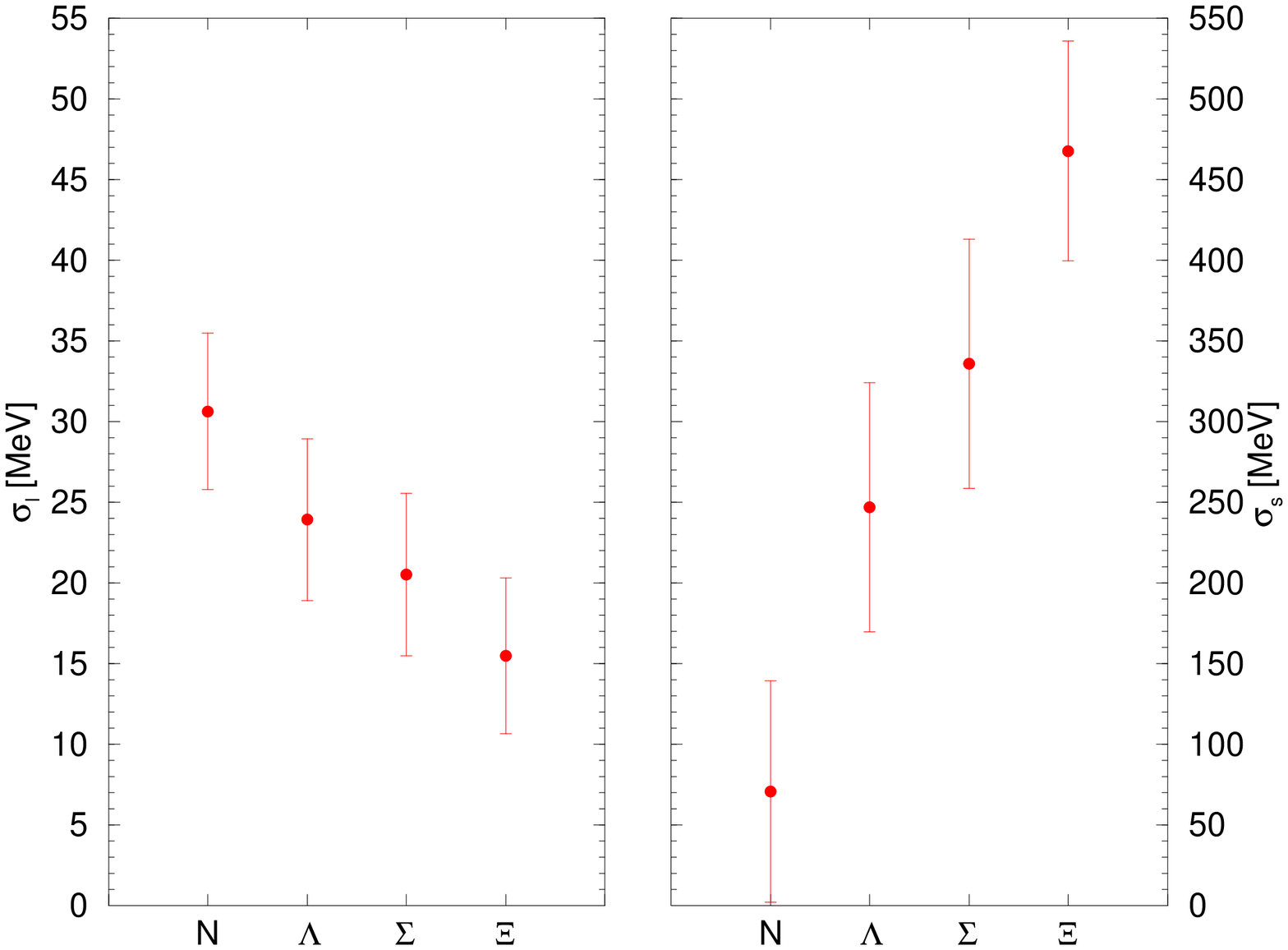}
   \end{center}
   \end{minipage}
   \caption{$y^{(H)\R*}$ (left plot) and $\sigma_l^{(H)*}$ and $\sigma_s^{(H)*}$
            (right plots) both at the physical point for
            $H = N$, $\Lambda$, $\Sigma$, $\Xi$ }
   \label{yN+b5p50_sigl+sigs_Xpi+XNscale}
\end{figure}

These results are illustrated in the left plot of 
Fig.~\ref{yN+b5p50_sigl+sigs_Xpi+XNscale} for $y^{(H)\R*}$
and in the right plots $\sigma_l^{(H)*}$, $\sigma_s^{(H)*}$ both against
$H = N$, $\Lambda$, $\Sigma$ and $\Xi$. While the data for $\kappa_0 = 0.12090$
is more complete than for $\kappa_0 = 0.12092$ (cf.\ the plots in
Fig.~\ref{mpsO2omNOpmSigOpmXiOo32_mNOomNOpmSigOpmXiOo3_32x64_lin})
and demonstrates linear behaviour, as the path starting at
$\kappa_0 = 0.12092$ is closer to the physical point
(cf.\ Fig.~\ref{b5p50_XNoXNkp12090_Xpi2oXpi2kp12090+mps2oXn2_2mpsK2-mps2oXn2})
we shall use these values as our final values.




In conclusion we have found that 
keeping the average quark mass constant gives very linear `fan' plots
from the flavour symmetric point down to the physical point.
This implies that an expansion in the quark mass from the flavour
symmetric point will give information about the physical point.
In this talk we have applied this to estimating the sigma terms
(both light and strange) of the nucleon octet. There has been no
use of a chiral perturbation expansion (indeed this is an opposite
expansion to the one used here, expanding about zero quark mass). 

Note that expansions about the
$SU(3)$ flavour line require consistency between many QCD observables,
here for example not only for the baryon octet under consideration here,
but also for the pseudoscalar octet, and PCAC and the ratio of
the light to strange quark mass.

Of course there are several more avenues to investigate.
Our approach here has been to emphasise linearity at
the expense (presently) of reaching exactly the physical point.
This can be addressed by interpolating between a small set
of constant $\overline{m}$ lines about the physical point.
Additionally the use of partial quenching will also help
to get closer to the physical pion mass. With more data,
a systematic investigation of quadratic quark mass terms
in the flavour expansion should be considered, to reduce the
systematic errors.


\section*{Acknowledgements}


The numerical configuration generation was performed using the
BQCD lattice QCD program, on the IBM
BlueGeneL at EPCC (Edinburgh, UK), the BlueGeneL and P at
NIC (J\"ulich, Germany), the SGI ICE 8200 at
HLRN (Berlin-Hannover, Germany) and the JSCC (Moscow, Russia).
We thank all institutions. The BlueGene codes were optimised using Bagel.
The Chroma software library, was used in the data analysis.
This work has been supported in part by the EU grants 227431
(Hadron Physics2), 238353 (ITN STRONGnet) and by the DFG under
contract SFB/TR 55 (Hadron Physics from Lattice QCD). JMZ is supported
by STFC grant ST/F009658/1.



\end{document}